\documentclass[final]{cvpr}

\usepackage{times}
\usepackage{epsfig}
\usepackage{graphicx}
\usepackage{amsmath}
\usepackage{amssymb}
\usepackage{booktabs}
\usepackage{multirow}

\usepackage[pagebackref=true,breaklinks=true,colorlinks,bookmarks=false]{hyperref}

\def\cvprPaperID{32} 
\def\confYear{CVPR 2021}

\begin{document}

\title{ Manipulation Detection in Satellite Images Using Vision Transformer }

\author{J{\'a}nos Horv{\'a}th, Sriram Baireddy, Hanxiang Hao, Daniel Mas Montserrat, Edward J. Delp\\
Video and Image Processing Laboratory (VIPER)\\ 
School of Electrical and Computer Engineering\\
Purdue University\\
West Lafayette, Indiana, USA
}

\maketitle

\begin{abstract}
A growing number of commercial satellite companies provide easily accessible satellite imagery.
Overhead imagery is used by numerous industries including agriculture, forestry, natural disaster analysis, and meteorology.
Satellite images, just as any other images, can be tampered with image manipulation tools.
Manipulation detection methods created for images captured by ``consumer cameras'' tend to fail when used on satellite images due to the differences in image sensors, image acquisition, and processing.
In this paper we propose an unsupervised technique that uses a Vision Transformer to detect spliced areas within satellite images.
We introduce a new dataset which includes manipulated satellite images that contain spliced objects.
We show that our proposed approach performs better than existing unsupervised splicing detection techniques.
\end{abstract}

\section{Introduction}
The exponentially growing number of commercial satellites orbiting the Earth generate an enormous amount of imagery.
A large variety of applications makes use of satellite imagery, including agricultural crop classification~\cite{gao_2020,russwurm_2019}, scene classification~\cite{amirabbas_2017,raiyani_2021}, wildlife monitoring~\cite{duporge_2020,guirado_2019}, forest characterization~\cite{helmer_2015,lee_2020}, meteorological analysis~\cite{lebedev_2019,pavuluri_2020}, infrastructure levels assessment, building localization~\cite{femin_2020,oshri_2018}, and soil moisture estimation~\cite{efremova_2018,foucras_2020}. 

\begin{figure}[t]
\centering
\includegraphics[width=\columnwidth]{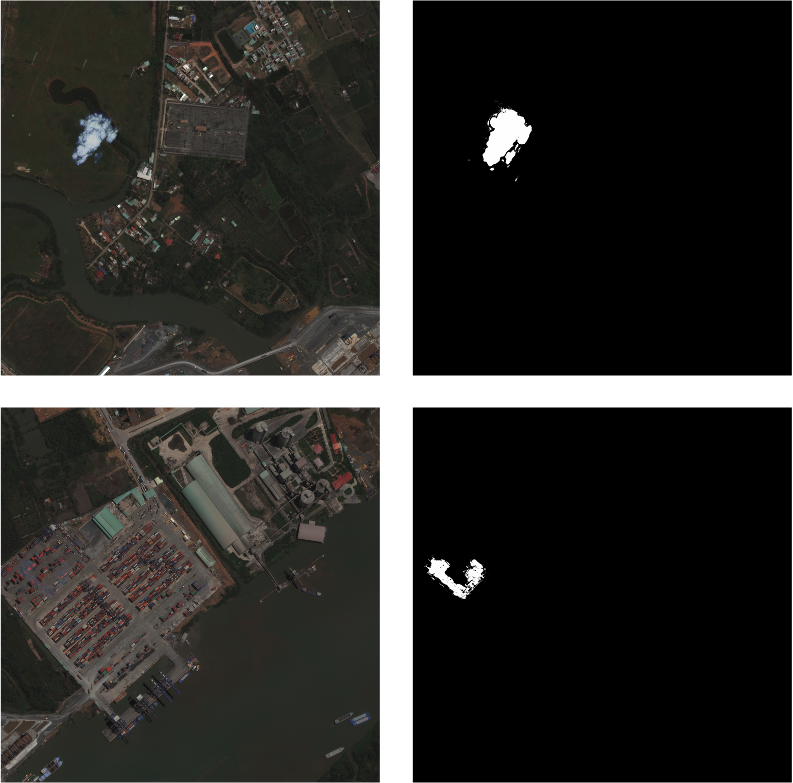}
\caption{Examples of satellite images: manipulated images (left) and their corresponding ground-truth of the spliced area (right).}
\label{fig:intro_pic_vt}
\end{figure}

Popular image editing tools, such as GIMP or Photoshop, can easily alter or manipulate satellite images. 
Figure \ref{fig:intro_pic_vt} shows some examples of manipulated satellite images.
Advances in machine learning have simplified the process of manipulating images and even creating highly-realistic ``fake'' images~\cite{horvath_2021,Zhou_2019}.
Several altered satellite images have been used to spread misinformation on the Internet.
Some examples include the Malaysian flight incident over Ukraine~\cite{kramer_2016}, the images of fake Chinese bridges~\cite{edwards_2019}, the Australian bushfires~\cite{georgina_2020}, and the Diwali Festival nighttime flyovers over India~\cite{byrd_2018}.

Several image manipulation methods have been proposed.
Some image manipulation techniques include splicing~\cite{carvalho_2013}, blending objects created by Generative Adversarial Networks (GANs)~\cite{christopher_2020}, and copy-move methods~\cite{verdoliva_2020}.
Various methods have been proposed to detect alterations in images captured by consumer cameras~\cite{bammey_2020,bayar_2016,cozzolino_2020,anderson_2011}. 
These techniques tend to fail in detecting alterations in satellite images due to the difference in the image types.
These differences include acquisition sensors, compression schemes, color channels, and post-processing operations like ortho-rectification.
Despite the growing number of techniques developed to detect manipulations in satellite imagery, it remains an open problem.

\thispagestyle{empty}

\section{Related Work}
\label{sec:related_work}

\begin{figure*}[!thpb]
\centering
\includegraphics[width=\textwidth]{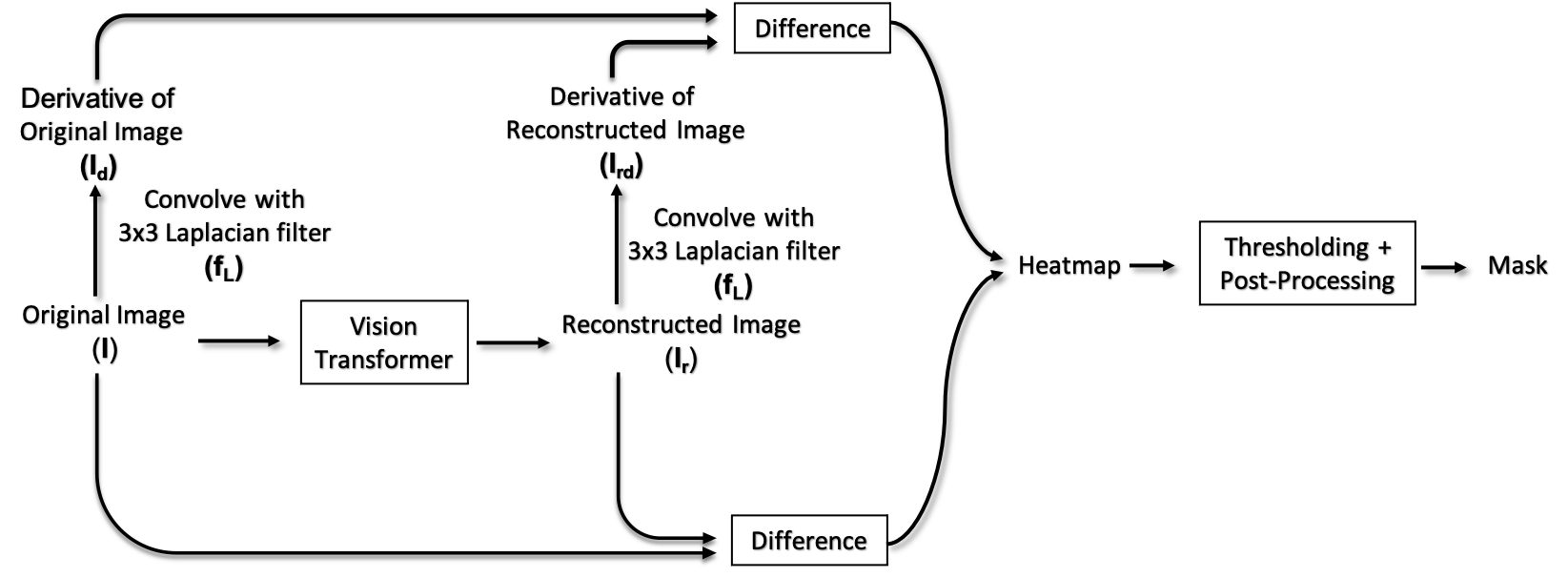}
\caption{Our proposed method for manipulation detection.}
\label{fig:method_overview_vt}
\end{figure*}

Several methods for detecting alterations in natural images have been described, including finding manipulated pixels in images by using neural networks with domain adaptation~\cite{cozzolino_2018}, using the Radon transform of resampled features and a deep learning classifier~\cite{bunk_2017}, and using saturation cues~\cite{mccloskey_2019}.
Other approaches focus on finding double-JPEG compression artifacts~\cite{barni_2010,wang_2016}
or detecting and localizing spliced objects in images using unsupervised approaches~\cite{bammey_2020,cozzolino_2015,cozzolino_2020}.
There has been work in detecting spliced regions by using local features in the image and differentiating between the original or splicing images using  expectation-maximization~\cite{cozzolino_2015}.
Cozzolino \etal~\cite{cozzolino_2020} developed a technique leveraging the fact that each camera model leaves a unique digital fingerprint, known as a ``noiseprint''.
Instead of finding general model-related artifacts, Bammey \etal~\cite{bammey_2020} designed a model focusing specifically on detecting camera demosaicing artifacts. 
A demosaicing artifact is a repeated pattern in the cycle of $2 \times 2$ pixel block. 
It is caused by the reconstruction of a full color image from the incomplete color samples in digital image process. 
This repeated pattern will be different between the manipulated region and original image area, which can be detected by the proposed DemosaicingNet~\cite{bammey_2020}. 
Most of these techniques will fail on satellite images due to the different acquisition process between images captured by consumer cameras and satellites.
These differences include different sensor technologies and post-processing steps such as the compression scheme and radiometric corrections.

Several methods for detecting manipulations in satellite images been proposed.
These detection techniques are based on more traditional techniques such as watermarking~\cite{ho_2005} or machine learning approaches that are unsupervised~\cite{horvath_2019,horvath_2020,sri_2018} and supervised~\cite{horvath_2021,bartusiak_2019}.
In~\cite{bartusiak_2019} a conditional GAN was used to localize spliced objects in satellite images.
In~\cite{horvath_2021} the Nested Attention U-Net for localizing spliced areas in overhead images was described.
The authors in~\cite{murtaza_2018} propose an authentication protocol for secure satellite image data transfer. 
While supervised methods tend to detect and localize spliced objects better than unsupervised approaches, they require both manipulated and original data during training.


For developing our method we consider the unsupervised scenario where no manipulated data is available during training.
The work introduced in~\cite{sri_2018} extracts and encodes patches from the input images into a lower dimensional latent space. 
This encoding is used by a one-class support vector machine (SVM) to determine if a patch contains a manipulation or not. 
Sat-SVDD, presented in~\cite{horvath_2019}, is a modified Support Vector Data Description (SVDD)~\cite{tax_2004}, meaning it is a one-class classifier that detects spliced objects in satellite images.
The Sat-SVDD input is patches extracted from the input image which are then encoded into a vector space within a hyper-sphere.
During inferencing, the patches whose vector representation are placed outside the hyper-sphere are considered ``altered'' patches. 
The authors in~\cite{horvath_2020} use a deep belief network (DBN)~\cite{hinton_2006_b} constructed by two layers of restricted Boltzmann machines (RBM)~\cite{smolensky_1986} following uniform distribution.
The Deep Belief Network is trained as an autoencoder which encodes and decodes the input patches.
The reconstruction error is then used to detect whether the patch contains alterations or not. 
The authors in work~\cite{mas_2020} use an ensemble of auto-regressive networks to detect forgeries in satellite images.
The ensemble predicts the probability of whether a pixel is manipulated or not.
This latest method performed better than the previously mentioned approaches.

In this paper we describe a splicing detection method using Vision Transformer~\cite{dosovitskiy_2020} and morphological filters~\cite{zhuang_1986}.
In the past Transformers were used mainly for natural language processing~\cite{wolf_2019}.
Recently, Transformer methods were developed for images~\cite{chen_2020,dosovitskiy_2020}
Image-GPT~\cite{chen_2020} was introduced last year for unsupervised low resolution image generation and image classification.
Image-GPT is an autoregressive network which aims to predict pixels without a complete understanding of the 2D input image structure.
This GPT model achieved high accuracy on the CIFAR10 dataset.
Another model developed for image classification is the Vision Transformer~\cite{dosovitskiy_2020}.
We will present an overview of Vision Transformer below in Section \ref{sec:method_vt}.
We will also introduce a morphological filter for binary images and use it in our post-processing step.

\section{Proposed Method}\label{sec:method_vt}


In this section we will describe an unsupervised splicing detection method for satellite images that uses Vision Transformer (ViT).
A block diagram of our proposed method is shown in Figure \ref{fig:method_overview_vt}.
In Section~\ref{sec:vit}, we will provide some background description of the ViT proposed by Dosovitskiy \etal~\cite{dosovitskiy_2020}. 
We extend ViT to an autoencoder-like structure for splicing detection in Section \ref{sec:vit-ours}.

\subsection{Vision Transformer}
\label{sec:vit}

\begin{figure*}[!thpb]
\centering
\includegraphics[width=\textwidth]{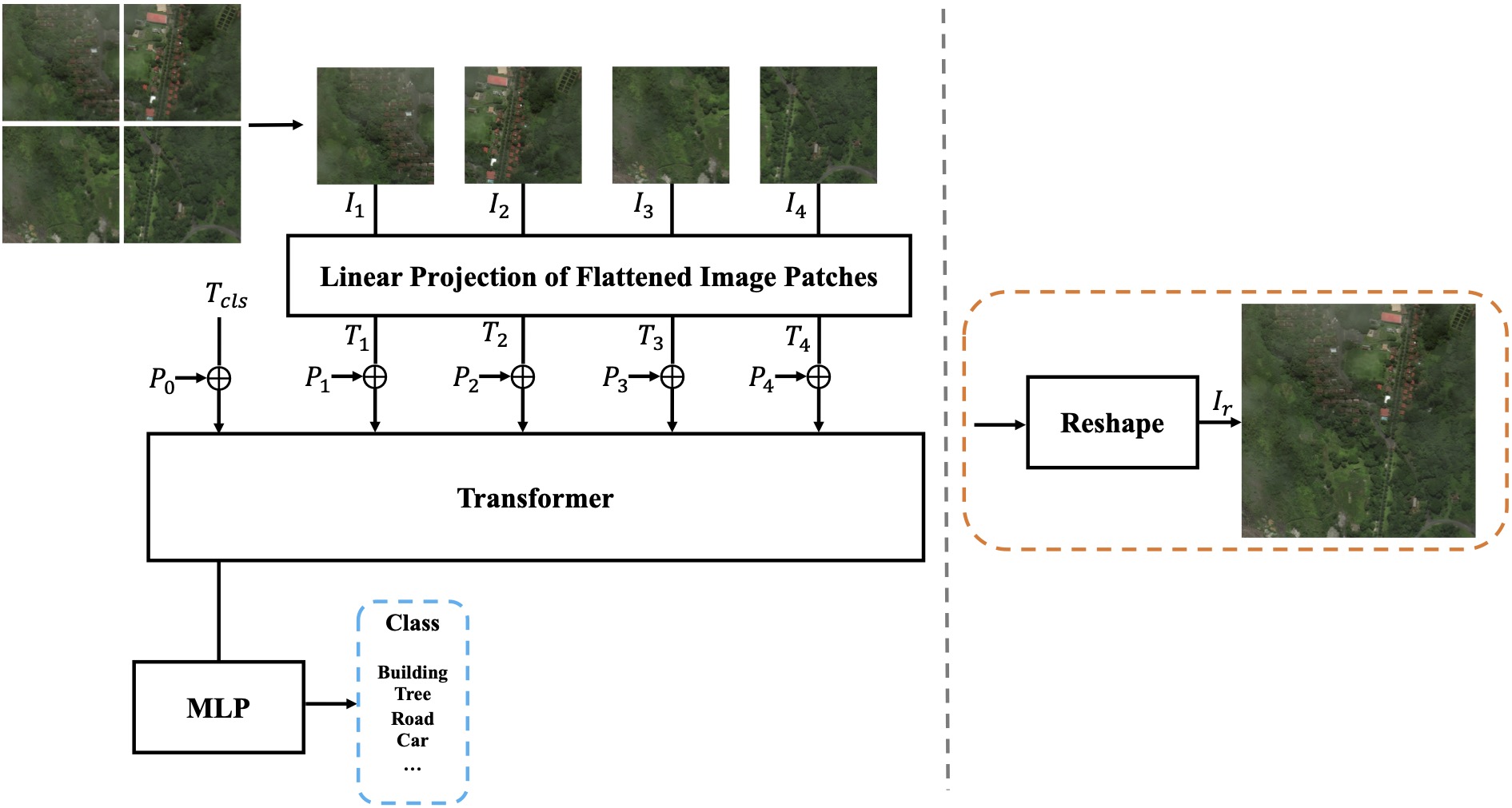}
\caption{Vision Transformer (ViT) block diagram (left) for image classification. Vision Transformer image reconstruction (right).}
\label{fig:vision_transformer_input_vector_vt}
\end{figure*}

As proposed in~\cite{dosovitskiy_2020}, ViT uses a Transformer model~\cite{vaswani_2017} to replace convolution layers for image classification tasks.
As shown in Figure \ref{fig:vision_transformer_input_vector_vt} (left), ViT takes image patches as input. 
In our experiments, the original image size is $128 \times 128$ and patch size is $64 \times 64$. 
To reduce the dimensionality of the input image patches, linear projection is used: $T_i = W \hat{I}_i$, where $W \in \mathbb{R}^{D \times N}$ is a learnable linear mapping function, $\hat{I}_i \in \mathbb{R}^{N}$ is the flattened $i$-th image patch, and $T_i \in \mathbb{R}^D$ is the $i$-th image token~\cite{dosovitskiy_2020} input to the Transformer.
As proposed in~\cite{dosovitskiy_2020}, we prepend a learnable classification token $T_0 \in \mathbb{R}^R$ to the aforementioned image tokens before inputting them to the Transformer model. 
Transformers use self-attention modules to add the long range information contained in all of the input tokens~\cite{vaswani_2017}.  
However, the self-attention module is invariant to the input token order.
To add positional information about the input patches to Transformer, a set of learnable positional embeddings~\cite{dosovitskiy_2020} are used: $P_i \in \mathbb{R}^D$ for $i \in \{0, 1, ...\}$.
These positional embeddings contain the unique position information for the different input tokens. 
After adding the positional embeddings to the input tokens, the position-aware tokens are provided to the Transformer. 
As proposed in~\cite{dosovitskiy_2020}, we only take the output from the classification token and pass it to a multi-layer perceptron (MLP) to output the probabilities of object classes.    

\subsection{Vision Transformer for Splicing Detection}
\label{sec:vit-ours}

Autoencoders have been successfully used for splicing detection~\cite{horvath_2019, horvath_2020,sri_2018}. 
They are trained to reconstruct images that do not contain manipulations. 
Then during testing, given a image with manipulated regions, the autoencoder will reconstruct the image using the information it learned from unmanipulated images.
We can compare the difference between the input image containing manipulation and the image reconstructed by the autoencoder. 
Since the autoencoder learned to model the image distribution of the original images, the reconstructed image will be different from the original image in the manipulated areas. 
Given this, we design a reconstruction approach as shown in the right side of Figure \ref{fig:vision_transformer_input_vector_vt} to replace the classification approach (\ie blue dot-line region in Figure \ref{fig:vision_transformer_input_vector_vt}).
We directly reshape the output from the MLP module to construct the output image $I_r$. 
To reduce the memory used by the self-attention module in the Transformer, we use the Linformer~\cite{wang_2020} to reduce the space complexity from the original Transformer used in~\cite{dosovitskiy_2020}. 
For our image reconstruction task, we use smoothed $L_1$ loss as following: 
\[
\resizebox{.99\hsize}{!}{$
\mathcal{L}_r(I, I_r) = \frac{1}{|I|} \sum_{i}^{|I|}
\begin{cases} 
  \frac{1}{2} (I(i) - I_r(i))^2 & |I(i) - I_r(i)| < 1 \\
  |I(i) - I_r(i)| - \frac{1}{2} & \text{Otherwise}
\end{cases}$
}
\]

At the inference stage, we input an image $\bold{I}$ into the trained Vision Transformer and obtain a reconstructed image $\bold{I_{r}}$ as seen in Figure \ref{fig:method_overview_vt}.
We also do a convolution for each channel of $\bold{I, I_{r}}$ with a $3\times3$ Laplacian filter $\bold{f_{L}}$, obtaining two more images $\bold{I_{d}, I_{rd}}$, where: $$\bold{I_{d}} = \bold{I} \circledast \bold{f_{L}}$$ and $$\bold{I_{rd}} = \bold{I_{r}} \circledast \bold{f_{L}}$$
The reason for using the Laplacian filter on the input and reconstructed images is that autoencoders have some difficulties when reconstructing the high frequency components of an image and the Laplacian filter acts an edge detector~\cite{lee_1986} that will highlight these high frequency components.
We construct a heatmap from the difference of $\bold{I, I_{r}}$ and from the difference of $\bold{I_{d}, I_{rd}}$ by averaging them.
Next, we threshold the heatmap to create a binary mask and use a post-processing stage to output a final mask that indicates the region detected as spliced. 


\subsection{Post-Processing with Morphological Filters}
\label{sec:post}

The post-processing consists of several morphological filters~\cite{haralick_1987}.
The goal of the post-processing is to decrease the number of false negatives and false positives.
This can be achieved by filling in holes and removing small objects in the binary mask.
There are efficient techniques which can fill holes in binary masks.
We use the function \textit{binary\_fill\_holes} from the SciPy library~\cite{virtanen_2020}.
After filling the holes in the mask, we remove the small objects.


\begin{figure}[!bhpb]
\centering
\includegraphics[width=0.6\columnwidth]{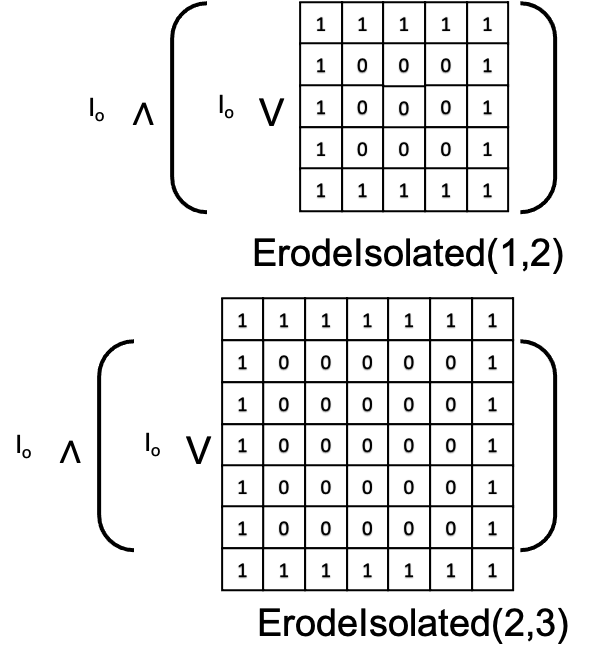}
\caption{Two example of $\bold{ErodeIsolated}$ morphological filter.}
\label{fig:morphological_filter_vt}
\end{figure}

For removing small objects, there are filters specifically designed for this task: erosion and opening~\cite{haralick_1987}.
While both of these filters remove small objects, they have the issue of also modifying larger objects.
Erosion decreases the area of large objects, while opening destroys the fine detail of the boundary~\cite{zhuang_1986}.
We propose \textbf{ErodeIsolated}, a morphological filter to erode smaller objects while leaving larger objects unchanged.
It has two parameters, $a$ and $b$, that are used to construct the structuring element:
$$\bold{f_{EI,a,b} = ErodeIsolated(a,b)}$$

Consider the first example shown in Figure \ref{fig:morphological_filter_vt}, where $a = 1$ and $b = 2$.
The resulting structural element has a square shape with a side length of $2b+1$ (in this case, $5$).
Within this structural element, we have an inner square with a side length of $2a+1$ (in this case, $3$).
We keep the values in the inner square at $0$ and the remaining values of the structural element at $1$.
By doing so, with \textbf{OR} ($\lor$) and \textbf{AND} ($\land$) operators as shown in Figure~\ref{fig:morphological_filter_vt}, we can remove small objects, while keeping the larger objects untouched.





\begin{figure}[!thpb]
\centering
\includegraphics[width=\columnwidth]{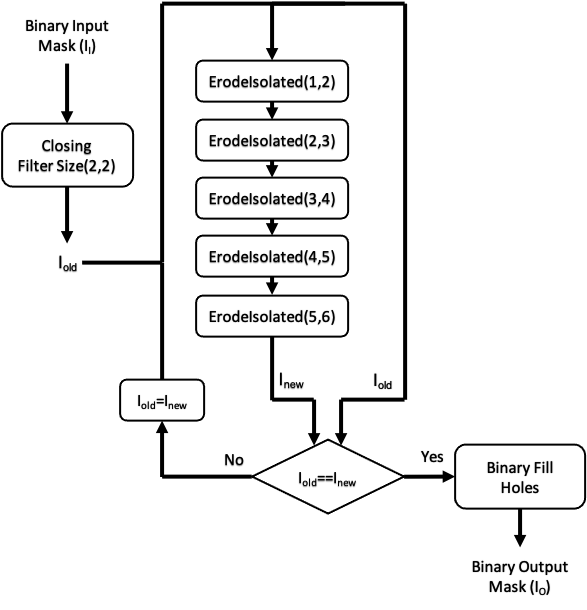}
\caption{Proposed post-processing method.}
\label{fig:postprocessing_vt}
\end{figure}

Our post-processing technique is shown in Figure \ref{fig:postprocessing_vt}. 
First, we use a closing operation with a very small structuring element, in order to ensure that all pixels in a large object are connected.
After that step we enter into a while loop.
In this while loop we use a series of ErodeIsolated filters with different structuring elements. 
The goal is to erode small objects while leaving the larger objects untouched. 
We exit the while loop when the series of filters no longer improves the binary image.
It is interesting to note that the while loop cannot iterate more than the number of ErodeIsolated filters inside the loop; in our case, this is no more than five.

\section{Experimental Results}


\begin{figure}[!thpb]
\centering
\includegraphics[width=\columnwidth]{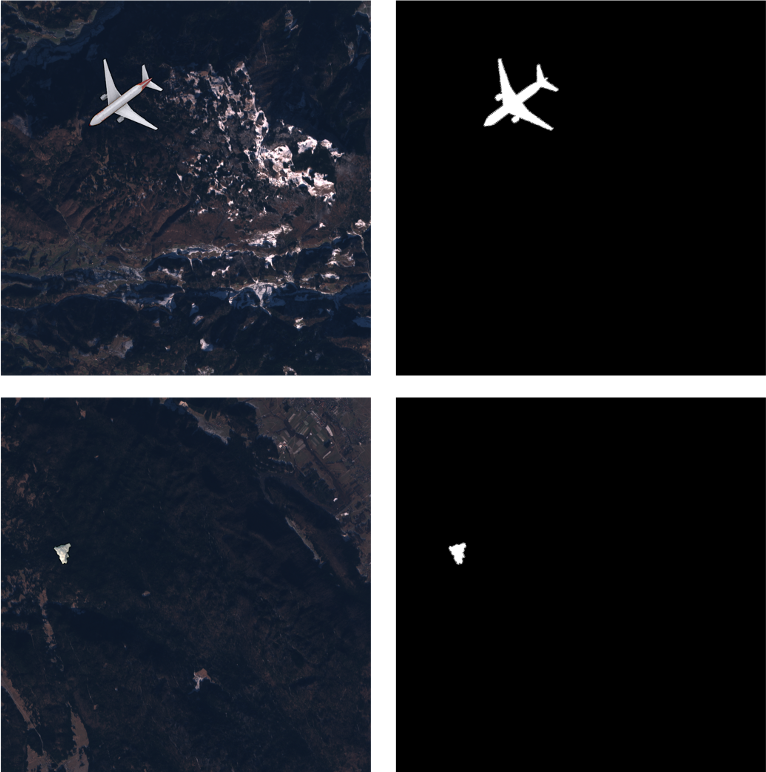}
\caption{$Dataset 1$: On the left are the manipulated images and on the right the corresponding ground-truth.}
\label{fig:dataset_vt_1}
\end{figure}

For evaluating the performance our proposed method we use two datasets in our experiments. 
The first dataset was introduced in~\cite{horvath_2020}, while the other dataset is new and will be discussed below.
The dataset described in~\cite{horvath_2020} is composed of satellite images of regions of Slovenia.
The images have dimensions of $1000\times1000$ and were captured by the Sentinel-2 satellite~\cite{Sentinel}.
We shall refer to this dataset as \emph{Dataset 1}.
We used 98 original images for training and 500 manipulated images with their corresponding ground-truth masks for testing.
Each manipulated image contains one spliced object randomly selected from nineteen different objects such as drones, planes, and clouds.
The objects are spliced into the images at different locations, rotation angles, and sizes including $16\times16$, $32\times32$, $64\times64$, $128\times128$, and $256\times256$ pixels.
Some examples from \emph{Dataset 1} are shown in Figure \ref{fig:dataset_vt_1}.

\begin{figure}[!thpb]
\centering
\includegraphics[width=\columnwidth]{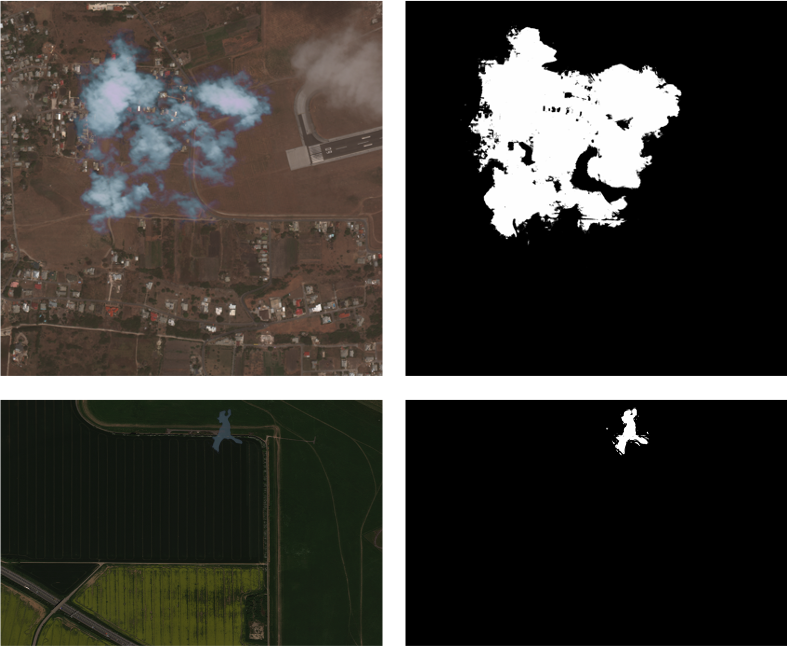}
\caption{$Dataset 2$: On the left are manipulated images and on the right the corresponding ground-truth.}
\label{fig:dataset_vt_2}
\end{figure}

We also constructed a new dataset, \emph{Dataset 2}, composed of satellite images captured by the WorldView-3 satellite from various locations such as coast, urban, and vegetation areas~\cite{xview_2}.
The resolution of these image varies from six to eight megapixels.
We used 28 images for training and 859 manipulated images with their corresponding ground-truth masks for testing.
Each manipulated image contains a spliced object extracted from images captured by a PlanetScope satellite~\cite{planetscope}. 
The objects were spliced into the WorldView-3 images using several steps including multiple blending functions.
Each image contains an object with sizes varying from several hundred pixels to several megapixels.
Some examples from \emph{Dataset 2} are shown in Figure \ref{fig:dataset_vt_2}.

We trained two Vision Transformers as autoencoders using the two datasets.
We assumed that we do not know prior information of the spliced objects; thus, we used only original images during training.
After training we tested our proposed technique on spliced images. 
We compared our method with previously introduced unsupervised splicing detection techniques, such as Splicebuster~\cite{cozzolino_2015}, NoisePrint~\cite{cozzolino_2020}, a Generative Ensemble of Gated PixelCNNs~\cite{mas_2020}, and DemosaicingNet~\cite{bammey_2020}.
The output of these techniques is a heatmap.
We thresholded the heatmaps in order to produce a binary mask as an output.

In order to evaluate the effectiveness of our proposed post-processing scheme, we compare different post-processing  methods.
First, as a baseline, we do not have any post-processing of the output mask; we refer to this method as ``Vision Transformer'' below.
The next post-processing approach consists of opening and closing operations~\cite{zhuang_1986}, both with a structuring element of ones in a $2 \times 2$ matrix, as well as \textit{binary\_fill\_holes} from the SciPy library~\cite{virtanen_2020}.
We refer to this as ``Vision Transformer with Post-Processing-V1''.
Finally, we use our proposed post-processing technique (shown in Figure \ref{fig:postprocessing_vt}), which we refer to  as ``Vision Transformer with Post-Processing-V2''.

\begin{figure*}[!thpb]
\centering
\includegraphics[width=0.9\textwidth]{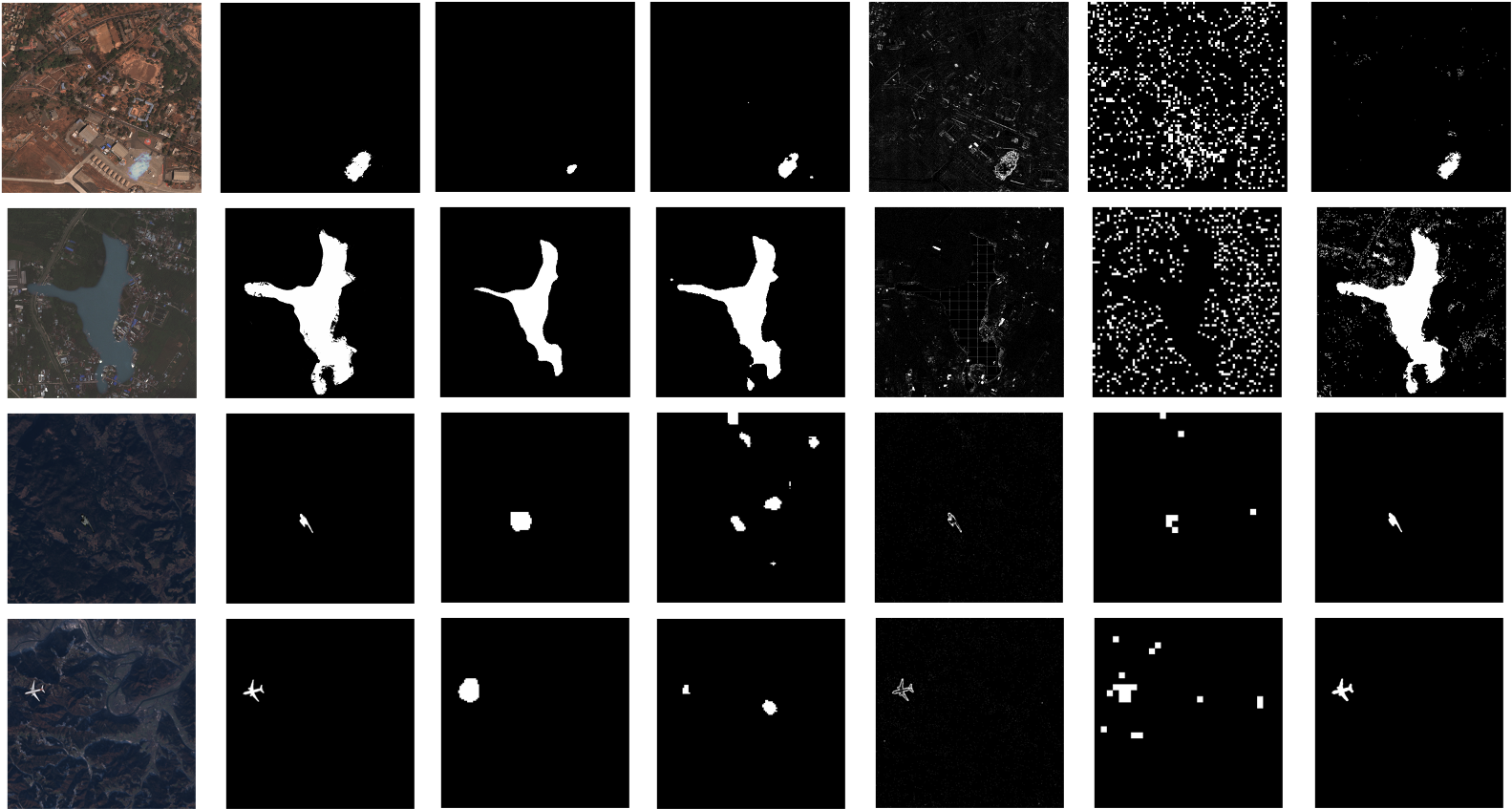}
\caption{The spliced image, its corresponding ground-truth mask, detection mask generated with Noiseprint, SpliceBuster,  Gated PixelCNN Ensemble, DemosaicingNet, Vision Transformer with Post-Processing-v2 }
\label{fig:results_vt}
\end{figure*}

\begin{table*}[!htb]
    \centering
	\caption {Results for \emph{Dataset 1}, where ``ViT'' stands for Vision Transformer and ``PP'' stands for post-processing} 
	\begin{tabular}{@{}lcccccccccc@{}}
	    \toprule
	    \textbf{Method} & F1$_{16}$    & F1$_{32}$  & F1$_{64}$ & F1$_{128}$ & F1$_{256}$ & JI$_{16}$ & JI$_{32}$  & JI$_{64}$ & JI$_{128}$ & JI$_{256}$  \\
	    \midrule
	    NoisePrint  & 0.000 & 0.001 & 0.066 & 0.148 & 0.174 & 0.000 & 0.001 & 0.042 & 0.096 & 0.111 \\
	    SpliceBuster & 0.001 & 0.012 & 0.094 & 0.360 & 0.504 & 0.000 & 0.006 & 0.049 & 0.219 & 0.337 \\
	    Gated PixelCNN Ensemble & 0.028 & 0.053 & 0.095 & 0.145 & 0.190 & 0.014 & 0.024 & 0.053 & 0.085 & 0.117 \\
	    DemosaicingNet & 0.004 & 0.013 & 0.057 & 0.120 & 0.181 & 0.002 & 0.007 & 0.029 & 0.064 & 0.100 \\
	    \midrule
	    ViT & 0.129 & 0.304 & 0.413 & 0.433 & 0.358 & 0.077 & 0.206 & 0.283 & 0.296 & 0.231 \\
	    ViT PP-v1 & 0.134 & 0.321 & 0.532 & 0.571 & 0.489 & 0.080 & 0.223 & 0.408 & 0.451 & 0.380 \\
	    \textbf{ViT PP-v2} & \textbf{0.215} & \textbf{0.411} & \textbf{0.614} & \textbf{0.694} & \textbf{0.672} & \textbf{0.140} & \textbf{0.302} & \textbf{0.493} & \textbf{0.582} & \textbf{0.587} \\
	    \bottomrule
	\end{tabular}
	\label{table:results_dt_1_vt}
\end{table*}

We used two evaluation metrics to characterize performance.
These two metrics measure the similarity between the generated masks and the corresponding ground-truth masks.
The first metric is the Dice Score, which is also known as the F1 score~\cite{dice_1945}.
A high F1 score, indicates that there is no problem with false positives or false negatives.
The F1 score is the harmonic mean of the Precision and Recall.
\begin{equation*}
F1 = \frac{Precision * Recall}{ 2* (Precision + Recall)}
\end{equation*}
or in an another form
\begin{equation*}
F1 = \frac{2*TP}{2*TP + FP + FN}
\end{equation*}
where $TP$ is the True Positive, $FP$ is the False Positive and $FN$ is the False Negative.
Precision is the ratio of True Positive to the sum of True Positive and False Positive. 
\begin{equation*}
Precision = \frac{TP}{TP + FP}
\end{equation*}
Recall is the proportion of positive scores that have been incorrectly predicted.
\begin{equation*}
Recall = \frac{TP}{TP + FN}
\end{equation*}
The second metric is the Jaccard Index, also known as intersection over union~\cite{LEVANDOWSKY_1971}.
\begin{equation*}
Jaccard \textit{ } Index \textit{ }(JI) = \frac{TP}{TP + FP + FN}
\end{equation*}

\section{Results}

\begin{figure*}
\centering
\includegraphics[width=0.95\textwidth]{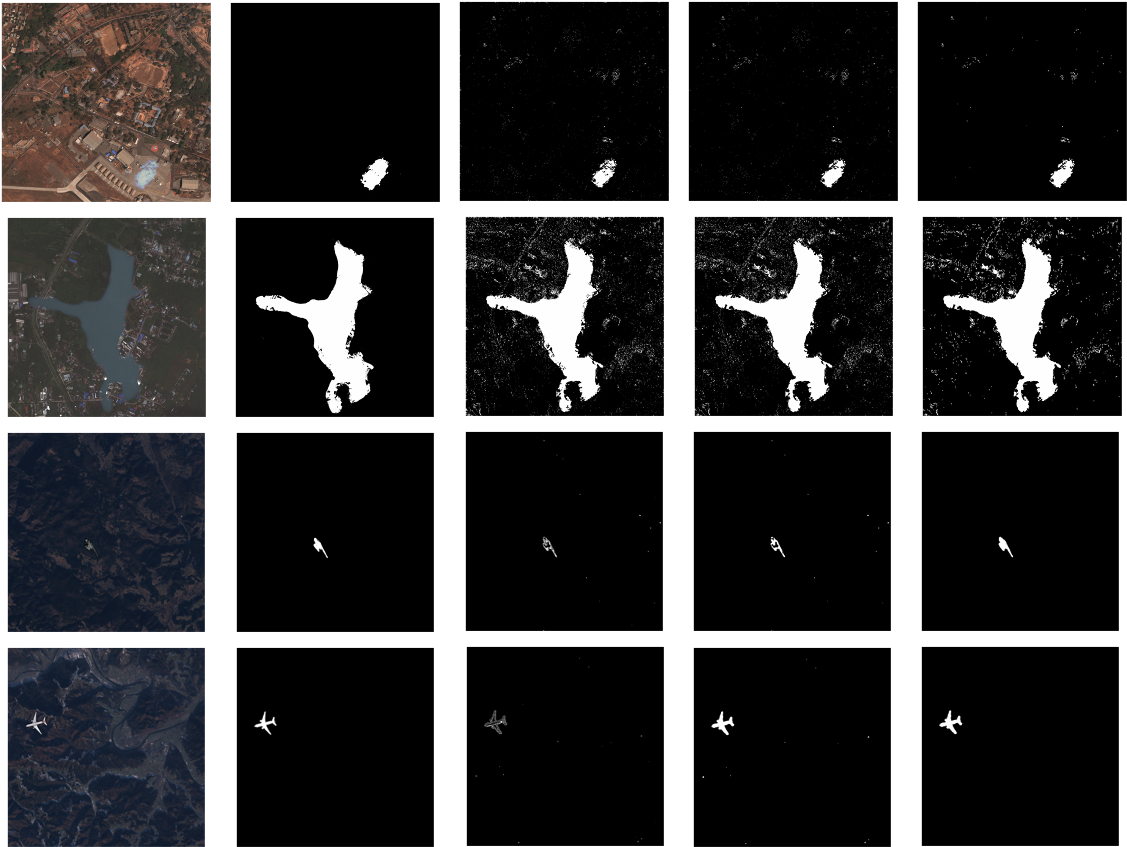}
\caption{The spliced image, its corresponding ground-truth mask, detection mask generated with Vision Transformer, Vision Transformer with Post-Processing-v1, and Vision Transformer with Post-Processing-v2 }
\label{fig:results_2_vt}
\end{figure*}

For \emph{Dataset 1} the spliced images can be grouped by the sizes of the spliced objects (\ie, $16\times16$, $32\times32$, $64\times64$, $128\times128$, and $256\times256$ pixels).
For \emph{Dataset 2} the spliced objects size varies from image to image, thus we cannot divide it into further groups.
Table \ref{table:results_dt_1_vt} shows the results for  \emph{Dataset 1}.
From this table we can see that all of the methods seem to detect larger spliced objects better than smaller objects.
We also see that the Vision Transformer generates a better splicing detection mask than NoisePrint, the Gated PixelCNN Ensemble, and DemosaicingNet.
Table \ref{table:results_dt_1_vt} shows in the last two rows that the Vision Transformer output mask can be improved by post-processing.
It also shows that if we use the \textbf{ErodeIsolated} filter we can produce a better splicing detection mask.
The difference between the two post-processing schemes is significant, as shown in Table \ref{table:results_dt_1_vt} and Table \ref{table:results_dt_2_vt}.


\begin{table}[!htb]
    \centering
	\caption {Results for \emph{Dataset 2}, where ``ViT'' stands for Vision Transformer and ``PP'' stands for post-processing} 
	\begin{tabular}{@{}lcc@{}}
	    \toprule
	    Method & F1 & JI  \\
	    \midrule
	    NoisePrint  & 0.066 & 0.037  \\
	    SpliceBuster & 0.337 & 0.202 \\
	    Gated PixelCNN Ensemble & 0.341 & 0.249  \\
	    DemosaicingNet & 0.022 & 0.011 \\
	    \midrule
	    ViT & 0.345 & 0.254 \\
	    ViT PP-v1 & 0.354 & 0.268  \\
	    \textbf{ViT PP-v2} & \textbf{0.364} & \textbf{0.275} \\
	    \bottomrule
	\end{tabular}
	\label{table:results_dt_2_vt}
\end{table}

Table \ref{table:results_dt_2_vt} shows the results using \emph{Dataset 2}.
From this table we can see that Vision Transformer generates a better splicing detection mask than NoisePrint and marginally better splicing detection mask than the Gated PixelCNN Ensemble and DemosaicingNet.
Both post-processing schemes output a better splicing detection mask and using ErodeIsolated is beneficial.

Figure \ref{fig:results_vt} presents two spliced images from each dataset with their corresponding ground-truth masks. 
It also shows the generated mask of our proposed method and the other techniques.
By visually inspecting these examples, we can see that NoisePrint is a good splicing detection method but fails to detect some details of the spliced objects and produces false positives. 
SpliceBuster does well on larger objects, but fails to localize smaller objects correctly. 
DemosaicingNet fails to detect the manipulated regions in both datasets, especially for the upper two cases as shown 
in Figure \ref{fig:results_vt}.
DemosaicingNet, which is designed for consumer cameras, is not able to provide an accurate detection. 
We conclude that Vision Transformer is better at detecting spliced objects in satellite images than NoisePrint, SpliceBuster, DemosaicingNet and the Gated PixelCNN Ensemble.
Using post-processing on the binary splicing detection mask improves the detection performance with respect to the Jaccard Index and Dice Score. 
The use of the \textbf{ErodeIsolated} morphological filter further improves the performance.
We show several examples in Figure \ref{fig:results_vt} to compare the different post processing schemes.

\section{Conclusion}

In this paper, we introduce an unsupervised splicing detection technique for detecting spliced objects in overhead images.
This technique uses Vision Transformer trained to localize manipulated areas.
We evaluated the performance of our approach on two datasets.
From the experiments the proposed method has better performance than previously introduced unsupervised splicing detection techniques.
In the future we plan to investigate other Transformer architectures to improve the performance.
We also plan to introduce more datasets for evaluating performance.

\section{Acknowledgment}

This material is based on research sponsored by the Defense Advanced Research Projects Agency (DARPA) and the Air Force Research Laboratory (AFRL) under agreement number FA8750-16-2-0173. 
The U.S. Government is authorized to reproduce and distribute reprints for Governmental purposes notwithstanding any copyright notation thereon. 
The views and conclusions contained herein are those of the authors and should not be interpreted as necessarily representing the official policies or endorsements, either expressed or implied, of DARPA or AFRL or the U.S. Government.

Address all correspondence to Edward J. Delp, ace@ecn.purdue.edu.

{\small
\bibliographystyle{ieee_fullname}
\bibliography{egbib}

\begin{thebibliography}{10}\itemsep=-1pt

\bibitem{bammey_2020}
Q. {Bammey}, R.~G. {von Gioi}, and J.~M. {Morel}.
\newblock {An Adaptive Neural Network for Unsupervised Mosaic Consistency
  Analysis in Image Forensics}.
\newblock {\em Proceedings of the IEEE Conference on Computer Vision and
  Pattern Recognition}, pages 14182--14192, June 2020.
\newblock Seattle, WA.

\bibitem{barni_2010}
M. Barni, A. Costanzo, and L. Sabatini.
\newblock { Identification of Cut\&Paste Tampering by Means of Double-{JPEG}
  Detection and Image Segmentation}.
\newblock {\em Proceedings of the IEEE International Symposium on Circuits and
  Systems}, pages 1687--1690, May 2010.
\newblock {Paris, France}.

\bibitem{bayar_2016}
B. Bayar and M.~C. Stamm.
\newblock {A Deep Learning Approach to Universal Image Manipulation Detection
  Using a New Convolutional Layer}.
\newblock {\em Proceedings of the ACM Workshop on Information Hiding and
  Multimedia Security}, page 5–10, June 2016.
\newblock Vigo, Galicia, Spain.

\bibitem{bunk_2017}
J. {Bunk}, J.~H. {Bappy}, T.~M. {Mohammed}, L. {Nataraj}, A. {Flenner}, B.~S.
  {Manjunath}, S. {Chandrasekaran}, A.~K. {Roy-Chowdhury}, and L. {Peterson}.
\newblock {Detection and Localization of Image Forgeries Using Resampling
  Features and Deep Learning}.
\newblock {\em Proceedings of the IEEE Conference on Computer Vision and
  Pattern Recognition}, pages 1881--1889, July 2017.
\newblock Honolulu, HI.

\bibitem{byrd_2018}
Deborah Byrd.
\newblock {Fake Image of Diwali Still Circulating}.
\newblock
  \url{https://earthsky.org/earth/fake-image-of-india-during-diwali-versus-the-real-thing}.

\bibitem{chen_2020}
M. Chen, A. Radford, R. Child, J. Wu, H. Jun, D. Luan, and I. Sutskever.
\newblock {Generative Pretraining From Pixels}.
\newblock {\em Proceedings of the International Conference on Machine
  Learning}, pages 1691--1703, July 2020.
\newblock {Virtual}.

\bibitem{cozzolino_2015}
D. {Cozzolino}, G. {Poggi}, and L. {Verdoliva}.
\newblock {Splicebuster: A New Blind Image Splicing Detector}.
\newblock {\em Proceedings of the IEEE International Workshop on Information
  Forensics and Security}, pages 1--6, November 2015.
\newblock {Rome, Italy}.

\bibitem{cozzolino_2018}
D. Cozzolino, J. Thies, A. Rossler, C. Riess, M. Niessner, and L. Verdoliva.
\newblock { ForensicTransfer: Weakly-Supervised Domain Adaptation for Forgery
  Detection }.
\newblock {\em arXiv preprint arXiv:1812.02510}, December 2018.

\bibitem{cozzolino_2020}
D. {Cozzolino} and L. {Verdoliva}.
\newblock {Noiseprint: A CNN-Based Camera Model Fingerprint}.
\newblock {\em IEEE Transactions on Information Forensics and Security},
  15:144--159, 2020.

\bibitem{amirabbas_2017}
A. Davari, V. Christlein, S. Vesal, A. Maier, and C. Riess.
\newblock {GMM Supervectors for Limited Training Data in Hyperspectral Remote
  Sensing Image Classification}.
\newblock {\em Proceedings of the International Conference on Computer Analysis
  of Images and Patterns}, pages 296--306, July 2017.
\newblock {Ystad, Sweden}.

\bibitem{carvalho_2013}
T. de Carvalho, C. Riess, E. Angelopoulou, H. Pedrini, and A. Rocha.
\newblock {Exposing digital image forgeries by illumination color
  classification}.
\newblock {\em IEEE Transactions on Information Forensics and Security},
  8(7):1182--1194, July 2013.

\bibitem{dice_1945}
L.~R. Dice.
\newblock { Measures of the Amount of Ecologic Association Between Species }.
\newblock {\em {Ecology}}, 26(3):297--302, July 1945.

\bibitem{dosovitskiy_2020}
A. Dosovitskiy, L. Beyer, A. Kolesnikov, D. Weissenborn, X. Zhai, T.
  Unterthiner, M. Dehghani, Matthias M., G. Heigold, S. Gelly, J. Uszkoreit,
  and N. Houlsby.
\newblock {An Image is Worth 16x16 Words: Transformers for Image Recognition at
  Scale}.
\newblock {\em arXiv preprint arXiv:2010.11929}, 2020.

\bibitem{duporge_2020}
I. Duporge, O Isupova, S. Reece, D. Macdonald, and T. Wang.
\newblock {Using Very High-Resolution Satellite Imagery and Deep Learning to
  Detect and Count African Elephants in Heterogeneous Landscapes}.
\newblock {\em Remote Sensing in Ecology and Conservation}, December 2020.

\bibitem{edwards_2019}
J. Edwards.
\newblock {China Uses GAN Technique to Tamper With Earth Images}.
\newblock
  \url{https://earthsky.org/earth/fake-image-of-india-during-diwali-versus-the-real-thing}.

\bibitem{efremova_2018}
N. Efremova, D. Zausaev, and G. Antipov.
\newblock {Prediction of Soil Moisture Content Based on Satellite Data and
  Sequence-to-Sequence Networks}.
\newblock {\em arXiv preprint arXiv:1907.03697}, June 2019.

\bibitem{femin_2020}
A. {Femin} and K.~S. {Biju}.
\newblock {Accurate Detection of Buildings from Satellite Images Using CNN}.
\newblock {\em Proceedings of the International Conference on Electrical,
  Communication, and Computer Engineering}, pages 1--5, June 2020.
\newblock {Istanbul, Turkey}.

\bibitem{foucras_2020}
M. {Foucras}, M. {Zribi}, and A. {Kallel}.
\newblock {Soil Moisture Estimation at 500m using Sentinel-1: Application to
  African Sites}.
\newblock {\em Proceedings of the International Conference on Advanced
  Technologies for Signal and Image Processing}, pages 1--5, September 2020.
\newblock Sousse, Tunisia.

\bibitem{gao_2020}
H. {Gao}, C. {Wang}, G. {Wang}, Q. {Li}, and J. {Zhu}.
\newblock {A New Crop Classification Method Based on the Time-Varying Feature
  Curves of Time Series Dual-Polarization Sentinel-1 Data Sets}.
\newblock {\em IEEE Geoscience and Remote Sensing Letters}, 17(7):1183--1187,
  October 2020.

\bibitem{guirado_2019}
E. Guirado, S. Tabik, M. L.~Rivas, D. Alcaraz-Segura, and F. Herrera.
\newblock {Whale Counting in Satellite and Aerial Images with Deep Learning}.
\newblock {\em Scientific Reports}, 9:14259--14259, October 2019.

\bibitem{haralick_1987}
R.~M. {Haralick}, S.~R. {Sternberg}, and X. {Zhuang}.
\newblock {Image Analysis Using Mathematical Morphology}.
\newblock {\em IEEE Transactions on Pattern Analysis and Machine Intelligence},
  9(4):532--550, April 1987.

\bibitem{helmer_2015}
E. Helmer, N.~R. Goodwin, V. Gond, C.~M. Souza, Jr., and G.~P. Asner.
\newblock {Characterizing Tropical Forests with Multispectral Imagery}.
\newblock In Prasad~S. Thenkaibail, editor, {\em Land Resources: Monitoring,
  Modeling and Mapping}, volume~2, pages 367--396. CRC Press, {Boca Raton,
  Florida}, 2015.

\bibitem{hinton_2006_b}
G.~E. Hinton, S. Osindero, and Yee-Whye Teh.
\newblock { A Fast Learning Algorithm for Deep Belief Nets }.
\newblock {\em Neural Computer}, 18(7), July 2006.

\bibitem{ho_2005}
A.~T.~S. {Ho} and W.~M. {Woon}.
\newblock { A Semi-Fragile Pinned Sine Transform Watermarking System for
  Content Authentication of Satellite Images }.
\newblock {\em Proceedings of the IEEE International Geoscience and Remote
  Sensing Symposium}, pages 1--4, July 2005.
\newblock {Seoul, South Korea}.

\bibitem{horvath_2019}
J. Horvath, D. Guera, S. K.~Yarlagadda, P. Bestagini, F. M.~Zhu, S. Tubaro, and
  E.~J. Delp.
\newblock { Anomaly-Based Manipulation Detection in Satellite Images}.
\newblock {\em Proceedings of the IEEE Conference on Computer Vision and
  Pattern Recognition Workshops}, pages 62--71, June 2019.
\newblock {Long Beach, CA}.

\bibitem{horvath_2021}
J. Horvath, D.~Mas Montserrat, and E.~J. Delp.
\newblock {Nested Attention U-Net: A Splicing Detection Method for Satellite
  Images}.
\newblock {\em Proceedings of the International Conference on Pattern
  Recognition}, pages 516--529, January 2021.
\newblock Virtual.

\bibitem{horvath_2020}
J. Horv{\'a}th, D.~Mas Montserrat, H. Hao, and E.~J. Delp.
\newblock { Manipulation Detection in Satellite Images Using Deep Belief
  Networks }.
\newblock {\em Proceedings of the IEEE Conference on Computer Vision and
  Pattern Recognition Workshops}, pages 2832--2840, June 2020.
\newblock {Seattle, WA}.

\bibitem{sri_2018}
S. K.~Yarlagadda, D. G\"{u}era, P. Bestagini, F. Zhu, S. Tubaro, and E.~J.
  Delp.
\newblock { Satellite Image Forgery Detection and Localization Using GAN and
  One-Class Classifier }.
\newblock {\em Proceedings of the IS\&T International Symposium on Electronic
  Imaging}, pages 214--1--214--9, February 2018.
\newblock {Burlingame, CA}.

\bibitem{kramer_2016}
A.~E. Kramer.
\newblock {Russian Images of Malaysia Airlines Flight 17 Were Altered, Report
  Finds}.
\newblock
  \url{https://www.nytimes.com/2016/07/16/world/europe/malaysia-airlines-flight-17-russia.html}.

\bibitem{planetscope}
Planet Labs.
\newblock {Planet Scope}.
\newblock \url{https://sentinel.esa.int/web/sentinel/missions}.

\bibitem{lebedev_2019}
V. Lebedev, V. Ivashkin, I. Rudenko, A. Ganshin, A. Molchanov, S. Ovcharenko,
  R. Grokhovetskiy, I. Bushmarinov, and D. Solomentsev.
\newblock {Precipitation Nowcasting with Satellite Imagery}.
\newblock {\em Proceedings of the ACM SIGKDD International Conference on
  Knowledge Discovery \& Data Mining}, pages 2680--2688, August 2019.
\newblock {Anchorage, AK}.

\bibitem{lee_1986}
J.S.J. Lee, L.G. Shapiro, and R.~M. Haralick.
\newblock {Morphologic Edge Detection}.
\newblock {\em Proceedings of the International Conference on Pattern
  Recognition}, pages 369--373, October 1986.
\newblock Paris, France.

\bibitem{lee_2020}
J.~H. {Lee}, J.~T.~S. {Sumantyo}, M.~M. {Waqar}, and J.~H. {Kim}.
\newblock {Analysis of Forest Loss by Sentinel-1 SAR Time Series}.
\newblock {\em Proceedings of the International Conference on Information and
  Communication Technology Convergence}, pages 182--184, October 2020.
\newblock Jeju, South Korea.

\bibitem{LEVANDOWSKY_1971}
M. Levandowsky and D. Winter.
\newblock { Distance Between Sets }.
\newblock {\em Nature}, 234:34--35, November 1971.

\bibitem{mccloskey_2019}
S. {McCloskey} and M. {Albright}.
\newblock { Detecting GAN-Generated Imagery Using Saturation Cues }.
\newblock {\em Proceedings of the IEEE International Conference on Image
  Processing}, pages 4584--4588, September 2019.
\newblock {Taipei, Taiwan}.

\bibitem{mas_2020}
D.~Mas Montserrat, J. Horvath, S.~K. Yarlagadda, F. Zhu, and E.~J. Delp.
\newblock { Generative Autoregressive Ensembles for Satellite Imagery
  Manipulation Detection }.
\newblock {\em Proceedings of the IEEE International Workshop on Information
  Forensics and Security}, pages 1--6, December 2020.
\newblock {Virtual}.

\bibitem{murtaza_2018}
A. Murtaza and L. Jianwei.
\newblock {A Simple, Secure and Efficient Authentication Protocol for Real-Time
  Earth Observation Through Satellite}.
\newblock {\em Proceedings of the International Bhurban Conference on Applied
  Sciences and Technology}, pages 822--830, January 2018.
\newblock Islamabad, Pakistan.

\bibitem{oshri_2018}
B. Oshri, A. Hu, P. Adelson, X. Chen, P. Dupas, J. Weinstein, M. Burke, D.
  Lobell, and S. Ermon.
\newblock {Infrastructure Quality Assessment in Africa Using Satellite Imagery
  and Deep Learning}.
\newblock {\em Proceedings of the International Conference on Knowledge
  Discovery \& Data Mining}, pages 616--625, August 2018.
\newblock {London, United Kingdom}.

\bibitem{pavuluri_2020}
B.~L. {Pavuluri}, R.~S. {Vejendla}, P. {Jithendra}, T. {Deepika}, and S.
  {Bano}.
\newblock {Forecasting Meteorological Analysis Using Machine Learning
  Algorithms}.
\newblock {\em Proceedings of the International Conference on Smart Electronics
  and Communication}, pages 456--461, September 2020.
\newblock {Tiruchirappalli, India}.

\bibitem{Sentinel}
Sentinel Program.
\newblock {Sentinel Missions}.
\newblock \url{https://sentinel.esa.int/web/sentinel/missions}.

\bibitem{bartusiak_2019}
E. R.~Bartusiak, S. K.~Yarlagadda, D. G\"{u}era, F. M.~Zhu, P. Bestagini, S.
  Tubaro, and E. J.~Delp.
\newblock { Splicing Detection And Localization In Satellite Imagery Using
  Conditional GANs }.
\newblock {\em Proceedings of the IEEE International Conference on Multimedia
  Information Processing and Retrieval}, pages 91--96, March 2019.
\newblock {San Jose, CA}.

\bibitem{raiyani_2021}
K. Raiyani, T. Goncalves, L. Rato, P. Salgueiro, and J. Silva.
\newblock {Sentinel-2 Image Scene Classification: A Comparison between Sen2Cor
  and a Machine Learning Approach}.
\newblock {\em Remote Sensing}, 13(2):300, January 2021.

\bibitem{georgina_2020}
G. Rannard.
\newblock {Australia fires: Misleading maps and pictures go viral}.
\newblock \url{https://www.bbc.com/news/blogs-trending-51020564}.

\bibitem{christopher_2020}
C.~X. Ren, A. Ziemann, J. Theiler, and A.~M.S. Durieux.
\newblock {Deep Snow: Synthesizing Remote Sensing Imagery with Generative
  Adversarial Nets}.
\newblock {\em arXiv preprint arXiv:2005.08892}, May 2020.

\bibitem{anderson_2011}
A. Rocha, W. Scheirer, T. Boult, and S. Goldenstein.
\newblock {Vision of the Unseen: Current Trends and Challenges in Digital Image
  and Video Forensics}.
\newblock {\em ACM Computing Surveys}, 43(4):26:1--26:42, October 2011.

\bibitem{russwurm_2019}
M. Ru{\ss}wurm, S. Lef{\`{e}}vre, and M. K{\"{o}}rner.
\newblock {BreizhCrops: A Satellite Time Series Dataset for Crop Type
  Identification}.
\newblock {\em arXiv preprint arXiv:1905.11893}, May 2019.

\bibitem{smolensky_1986}
P. Smolensky.
\newblock {\em { Information Processing in Dynamical Systems: Foundations of
  Harmony Theory }}.
\newblock MIT Press, {Cambridge, MA}, 1986.

\bibitem{tax_2004}
D.~M.~J. Tax and R.~P.~W. Duin.
\newblock {Support Vector Data Description}.
\newblock {\em Machine Learning}, 54(1):45--66, January 2004.

\bibitem{xview_2}
Defense~Innovation Unit.
\newblock {xView2}.
\newblock \url{https://www.xview2.org}.

\bibitem{vaswani_2017}
A. Vaswani, N. Shazeer, N. Parmar, J. Uszkoreit, L. Jones, A.~N. Gomez, L.
  Kaiser, and I. Polosukhin.
\newblock {Attention is All You Need}.
\newblock {\em Proceedings of the International Conference on Neural
  Information Processing Systems}, pages 6000--6010, December 2017.
\newblock Long Beach, CA.

\bibitem{verdoliva_2020}
L. Verdoliva.
\newblock { Media Forensics and DeepFakes: an overview}.
\newblock {\em arXiv preprint arXiv:2001.06564}, January 2020.

\bibitem{virtanen_2020}
P. Virtanen, R. Gommers, T.~E. Oliphant, M. Haberland, T. Reddy, D. Cournapeau,
  E. Burovski, P. Peterson, W. Weckesser, J. Bright, S.~J. {van der Walt}, M.
  Brett, J. Wilson, K.~J. Millman, N. Mayorov, A.~R.~J. Nelson, E. Jones, R.
  Kern, E. Larson, C~J Carey, I. Polat, Y. Feng, E.~W. Moore, J. {VanderPlas},
  D. Laxalde, J. Perktold, R. Cimrman, I. Henriksen, E.~A. Quintero, C.~R.
  Harris, A.~M. Archibald, A.~H. Ribeiro, F. Pedregosa, P. {van Mulbregt}, and
  {SciPy 1.0 Contributors}.
\newblock {{SciPy} 1.0: Fundamental Algorithms for Scientific Computing in
  Python}.
\newblock {\em Nature Methods}, 17:261--272, February 2020.

\bibitem{wang_2016}
Q. Wang and R. Zhang.
\newblock {Double JPEG Compression Forensics Based on a Convolutional Neural
  Network}.
\newblock {\em EURASIP Journal on Information Security}, 2016(23), October
  2016.

\bibitem{wang_2020}
S. Wang, B.~Z. Li, M. Khabsa, H. Fang, and H. Ma.
\newblock { Linformer: Self-Attention with Linear Complexity}.
\newblock {\em arXiv preprint arXiv:2006.04768}, June 2020.

\bibitem{wolf_2019}
T. Wolf, L. Debut, V. Sanh, J. Chaumond, C. Delangue, A. Moi, P. Cistac, T.
  Rault, R. Louf, M. Funtowicz, and J. Brew.
\newblock {HuggingFace's Transformers: State-of-the-Art Natural Language
  Processing}.
\newblock {\em arXiv preprint arXiv:1910.03771}, June 2020.

\bibitem{Zhou_2019}
X. Zhou, S. Huang, B. Li, Y. Li, J. Li, and Z. Zhang.
\newblock {Text Guided Person Image Synthesis}.
\newblock {\em Proceedings of the IEEE Conference on Computer Vision and
  Pattern Recognition}, pages 3658--3667, June 2019.
\newblock {Long Beach, CA}.

\bibitem{zhuang_1986}
X. Zhuang and R.~M. Haralick.
\newblock { Morphological Structuring Element Decomposition }.
\newblock {\em Computer Vision, Graphics, and Image Processing}, 35:370--382,
  April 1986.

\end{thebibliography}
}

\end{document}